# Dispersion Polymerization in an Elastomeric Solvent

*Senthilkumar Duraivel[1], Reagan J. Dreiling[2], Tyler E. Ball[2], Harsha Koganti[1†], Jay Fu[1],*

*Ethan E. O'Banion[1], Brett P. Fors[2], Eric R. Dufresne[1,3*]*

[1]Department of Materials Science and Engineering, Cornell University, Ithaca, NY 14853, USA

[2]Department of Chemistry and Chemical Biology, Cornell University, Ithaca, NY 14853, USA

[3]Department of Physics, Cornell University, Ithaca, NY 14853, USA

KEYWORDS: polymerization-induced phase separation, dispersion polymerization, microparticles, composites, elastomers, controlled polymerization

## ABSTRACT

Polymerization-induced phase separation (PIPS) provides a powerful route to generate structured polymeric materials by coupling chemical conversion with thermodynamic demixing. PIPS in liquid-state systems underlies dispersion polymerization, serving as a cornerstone technique for microparticle production, yet is constrained by solvent compatibility and limited range of morphologies. In contrast, PIPS within solid or polymeric media produces kinetically arrested, co-continuous structures with limited tunability. Here, we establish an elastically mediated PIPS regime that bridges these two limits by conducting controlled polymerization within a deformable elastomeric network. This approach, termed Dispersion Polymerization in an Elastomeric Solvent (DiPolES), serves as a solid-state analogue of dispersion polymerization in which an elastomeric network simultaneously serves as solvent and physical stabilizer. Using photoiniferter-mediated polymerization of methyl methacrylate (MMA) within poly(dimethyl siloxane) (PDMS) elastomeric solvent, DiPolES enables robust fabrication of elastomeric composites containing uniform PMMA microparticles with tunable size (0.85 to 3 μm) and shape (spheroidal and ellipsoidal). The strategy is generalizable beyond the PDMS/MMA system and is applicable to diverse monomers, such as acrylonitrile and 2-vinyl pyridine, which can be extracted from the elastomeric solvent, enabling high-yield production of microparticles. Real-time imaging and compositional analysis reveal that particle formation proceeds through rapid nucleation at low monomer conversion, followed by growth accompanied by cavitation of the surrounding network. Monomer loading governs the particle size, while solvent elasticity modulates the transition from isolated uniform spheroids to heterogeneous clusters. Interestingly, applying uniaxial strain during DiPolES enables production of ellipsoidal particles without any post-processing. By establishing elastic confinement as a fundamental parameter for directing PIPS, DiPolES provides a general framework for designing multiphase composite materials with tunable properties.

## 1. Introduction

Phase separation is a central organizing principle in materials science and provides a general route by which molecular interactions are translated into microstructures. When a multicomponent system becomes thermodynamically unstable, the homogenous mixture spontaneously phase separates into chemically distinct domains with characteristic dimensions from nano- to macro-scales.[1–4] Such self-organization leverages intrinsic thermodynamic forces rather than external templating and therefore offers an efficient pathway for the bulk fabrication of microstructured materials. In polymer systems, this principle is frequently realized through polymerization-induced phase separation (PIPS).[5–7] During PIPS, progressive conversion of monomer into polymer results in reduced configurational entropy, which drives demixing into segregated domains. By coupling thermodynamic demixing with sophisticated polymerization schemes, PIPS provides a versatile platform for generating multiphase composite materials with diverse structural morphologies from a plethora of polymeric materials.[8–11]

In liquid-state dispersed systems, PIPS underlies classical dispersion and precipitation polymerizations, where growing polymer chains lose solubility, nucleate droplets, and ultimately form stabilized particles.[12–14] These methods constitute the dominant strategy for producing polymer microparticles, which serves as scaffolds in many modern technologies, including biomedicine, photonics, catalysis, and environmental remediation.[15–19] Despite these successes, dispersion polymerization is fundamentally and practically limited: high monomer loading limits the achievable particle yield due to aggregation,[13,14] formulation-specific stabilizer requirements limit generality and complicates purification due to potential adsorption or covalent grafting to the polymer,[20] and complex solvent–stabilizer–polymer interactions constrain the range of monomers and functional additives that can be incorporated. Furthermore, the lack of physical stabilization and minimization of interfacial energy limits particle morphologies to be spherical[21], and multistep postprocessing strategies are required to achieve anisotropic particle shapes [cite].

Overcoming these limitations arising from complex interfacial chemistries requires a conceptual shift in how dispersed polymer phases are generated and stabilized. A promising strategy is to replace the liquid solvent with an elastomeric polymer network, where the elastic forces of the network govern the stability and morphology of phase-separated domains. For instance, PIPS in entangled glassy polymers[22,23] and heavily crosslinked rubber matrices[24,25] has revealed that demixed phases can kinetically arrest into spheroidal particles or topologically continuous structures, and that nascent phase-separated domains vitrify and remain stable over long periods of time. While polymeric systems eliminate the need for surfactants, the resulting morphologies are typically heterogeneous, confined to limited size range, and difficult to tune systematically as the length scales are dictated by vitrification or gelation rather than controlled nucleation and growth.[26,27] Though the mechanics of the solid medium has been underexplored as a variable to tune the morphology of controlled particles in PIPS, we recently demonstrated that mechanics of lightly-crosslinked elastomers could be used to precisely control the size and shape of microphases produced via thermally-induced phase separation.[28–30] Therefore, we envision that a lightly crosslinked elastomeric network would be an ideal solvent medium for PIPS, wherein mechanical stresses act as an independent thermodynamic variable, enabling phase separation to be directed not only by interfacial energetics but also by stress fields within the medium.

Building upon these insights, we introduce an elastically mediated PIPS paradigm by conducting photocontrolled polymerization within an elastomeric solvent. This strategy, termed Dispersion Polymerization in an Elastomeric Solvent (DiPolES), creates a solid-state analogue of dispersion polymerization in which the elastic network functions simultaneously as chemical solvent and physical stabilizer. Employing photoiniferter-mediated polymerization of various monomers within an elastomeric solvent, we demonstrate that DiPolES enables robust fabrication of size- and shape-controlled polymeric microparticles. This polymeric phase can either remain embedded in the elastomeric solvent to be utilized as composites or can be cleanly extracted by depolymerizing the elastomeric network to generate microparticles. Real-time imaging and compositional analysis reveal that particle formation proceeds through a rapid nucleation stage at low monomer conversion, followed by growth accompanied by cavitation of the surrounding network. The resulting particles exhibit narrow size distributions and high yield, with size tunability governed by monomer loading rather than solvent compatibility. Beyond stabilization, the elastomeric solvent provides a mechanical handle on the particle morphology, with stiffer matrices resulting in clustered particles and stretched matrices creating ellipsoidal particles. Overall, these results establish DiPolES as a modular and tunable platform for fabricating multiphase elastomeric composites and extract microparticles, by using an elastomeric solvent to circumvent the inherent limitations of liquid-state dispersion polymerization and uncontrolled solid-state PIPS. By coupling controlled polymerization with elastic stabilization, DiPolES expands the accessible design space of hierarchically organized, mechanically coupled composites and microparticle fabrication.

## 2. Results and Discussion

### 2.1 Dispersion polymerization in an elastomeric solvent (DiPolES) Yields Uniform Microparticles from a Diverse Monomer Palette

The thermodynamics underlying dispersion polymerization can be visualized with a simplified phase diagram, wherein we consider only three components: monomer, polymer, and solvent (Fig. 1A). A full thermodynamic description would account for the distribution of polymer chain lengths. Initially, the reaction mixture - composed of monomer and solvent - exists as a homogeneous single phase starting on the monomer-solvent axis. As polymerization proceeds, the system traverses along the green arrow, parallel to the monomer-polymer axis. As chains grow, the entropic incentive to mixing vanishes, favoring separation into polymer-rich and solvent-rich phases, whose compositions are denoted by the black-dashed tie lines.

DiPolES would leverage the same thermodynamic principles but prevent aggregation of the phase-separated domains by utilizing a lightly crosslinked elastomeric solvent (Fig. 1B). Inspired by previous studies of LLPS within elastomers[31,32] and in the broader context of mechanically constrained polymerization systems [cite], we expected that objects significantly larger than the mesh size of the elastomer could be trapped, which would prevent aggregation of phase-separated domains and eliminate the need for surfactant stabilization. We expect crosslinks to have little effect on polymerization kinetics, since small objects like monomers or short polymer chains are free to diffuse. A schematic of our strategy is outlined in Figure 1C. Crosslinked polydimethylsiloxane (PDMS), functioning as an elastomeric solvent, is swollen with a defined amount of polymerization mixture, consisting of a radically polymerizable monomer (methyl methacrylate, acrylonitrile, or 2-vinyl pyridine) and iniferter/chain transfer agent (CTA) (2-cyano-2-propyl dodecyl trithiocarbonate). The sample is then irradiated with either 370 nm or 456 nm light to initiate photoiniferter-mediated polymerization. After running the reaction close to completion (Fig. S1), the morphology of the new polymer phase can be characterized using optical microscopy.

We began our investigation by swelling crosslinked PDMS possessing a Young's modulus ($E$) of 100 kPa, with methyl methacrylate (MMA) and trithiocarbonate CTA, targeting a monomer loading ($C_{liq}$) of 30% (see *Supporting Information* for full experimental details). After 4 h of irradiation with 370 nm light, stable PMMA particles were observed by optical microscopy (see *Supporting Information* for imaging details) (Fig. 1C). Monomer conversion to polymer was measured by swelling the PDMS elastomer in $CDCl_3$ for 16 hours; after 4 hours of irradiation, we observed ~80-90% conversion of MMA to polymer. Similarly, molecular weight distributions were obtained by swelling the sample in tetrahydrofuran (THF) to remove polymer chains and running the sample on SEC. A typical polymerization yielded $M_n$ = 6.8 kg/mol and $Đ$ = 1.95, indicating that control over polymerization is lost over the course of the reaction. We expect DiPolES to produce particles with flexible polymer chemistries, provided the monomer of choice can sufficiently swell the crosslinked PDMS. To investigate the chemical versatility of DiPolES, we similarly swelled PDMS ($E$ = 100 kPa) with acrylonitrile (AN) and 2-vinyl pyridine (2-VP). Polymerization of both AN ($C_{liq}$ = 29 wt%) and 2-VP ($C_{liq}$ = 39 wt%) yielded stable particles, as confirmed by brightfield and fluorescence microscopy (Fig. 1C). Due to the insolubility of polyacrylonitrile (PAN) in solvents that will swell the PDMS elastomer, we were unable to obtain conversions or molecular weights for this sample. Polymerization of 2-VP ran to 12% conversion to yield polymers with $M_n$ = 19.8 kg/mol and $Đ$ = 1.69. Future optimization will enable efficient polymerization of these monomers. Whereas a low monomer loading (10% w/v) and distinct solvents, stabilizers, and initiators are required to produce these chemically distinct particles via conventional dispersion polymerization, DiPolES produces them all using a single set of reaction conditions, albeit with varying irradiation times.[33–35] Particles produced by DiPolES were extracted from the elastomeric solvent using tetrabutylammonium difluorotriphenylsilicate (TBAT)-catalyzed depolymerization of PDMS (see *Supporting Information* for details) (Fig. S2).[36] Further expansion of the monomer scope under DiPolES can be achieved through the use of alternative elastomeric solvents, which we will explore in future studies.

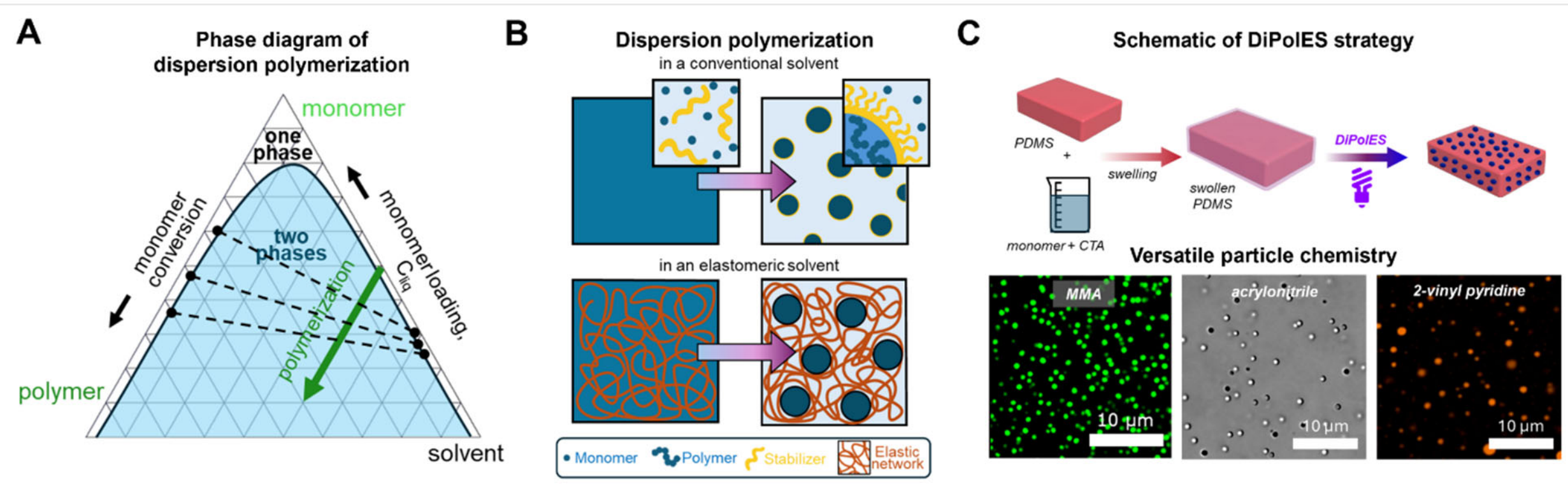

**Figure 1.** Dispersion polymerization in an elastomeric solvent (DiPolES) to produce particles of homogeneous size with versatile monomers. (A) Qualitative phase diagram. (B) Schematic comparison of dispersion polymerizations in typical solvent (top) versus an elastomeric solvent (bottom). (C) Brightfield and fluorescence micrographs of microparticles formed from MMA, AN, and 2-VP (left to right). Typical reaction conditions: 100 equiv monomer, 1 equiv trithiocarbonate, PDMS ($E$ = 100 kPa), 16 h dark swelling, 4 h 370 or 456 nm h*v*, $N_2$ atmosphere.

## 2.2 Real-time imaging reveals particle nucleation at low conversions during DiPolES

To elucidate the mechanism of particle formation in DiPolES, we monitored the reaction in real-time using brightfield microscopy (Fig. 2A). PDMS swollen with MMA/CTA ($C_{liq}$ = 40%) was sandwiched between two coverslips and sealed in an inert atmosphere with epoxy putty. The sample was irradiated with 370 nm light to initiate polymerization and imaged in 5 second intervals through an optical filter that blocks the wavelength range corresponding to the photopolymerization light source. Time-lapse imaging of the photopolymerization process reveals that discrete particles nucleate uniformly within 2 minutes of irradiation (Fig. 2B). Particle nucleation and growth was systematically quantified by measuring the number density and volume fraction of particles formed at different photopolymerization times during DiPolES (Figure 2C). After a short burst of nucleation over the first 20 minutes of irradiation, the number of particles remains essentially constant, while the particle volume fraction continues to grow steadily as irradiation continues.

The role of polymerization kinetics on particle nucleation was investigated by swelling PDMS ($E$ = 100 kPa) with MMA/CTA ($C_{liq}$ = 40%) and characterizing monomer conversion and molecular weight ($M_n$) evolution of PMMA over time using proton nuclear magnetic resonance spectroscopy ($^1$H NMR) and gel permeation chromatography (GPC), respectively(Fig. S3, S4). At early polymerization times we observe good matching between theoretical and experimental $M_n$ and low dispersity ($Đ$) but note a loss of control as irradiation time increases. Furthermore, we note the appearance of a high molecular weight tail at increased irradiation times, indicating uncontrolled polymerization. We theorize that this loss of control is  due to increased turbidity leading to decreased light penetration as the sample is irradiated. Interestingly, particle morphologies at $C_{liq}$ = 40% were independent of the irradiation wavelength (370 nm vs. 456 nm) or light intensity (50% vs. 100%) used to initiate polymerization (Fig. S5). Relating polymerization kinetics to calculated volume fractions and particle densities revealed that the point at which particle nucleation ends (~30 minutes) corresponded to 20% monomer conversion (Fig. 2C, left, Fig. S6A). After 45 minutes, particle growth begins to plateau, even though monomer conversion continues to grow rapidly until 60 minutes (Fig. 2C, right, Fig. S6B). This observation suggests that the initially nucleated “particles” are not fully polymerized solids, but rather mixtures of monomer and polymer (Fig. S7). Thus, the replacement of a liquid solvent with an elastomeric one preserves many of the characteristic features of dispersion polymerization in nonpolar solvents, including nucleation followed by particle growth.[37,38]

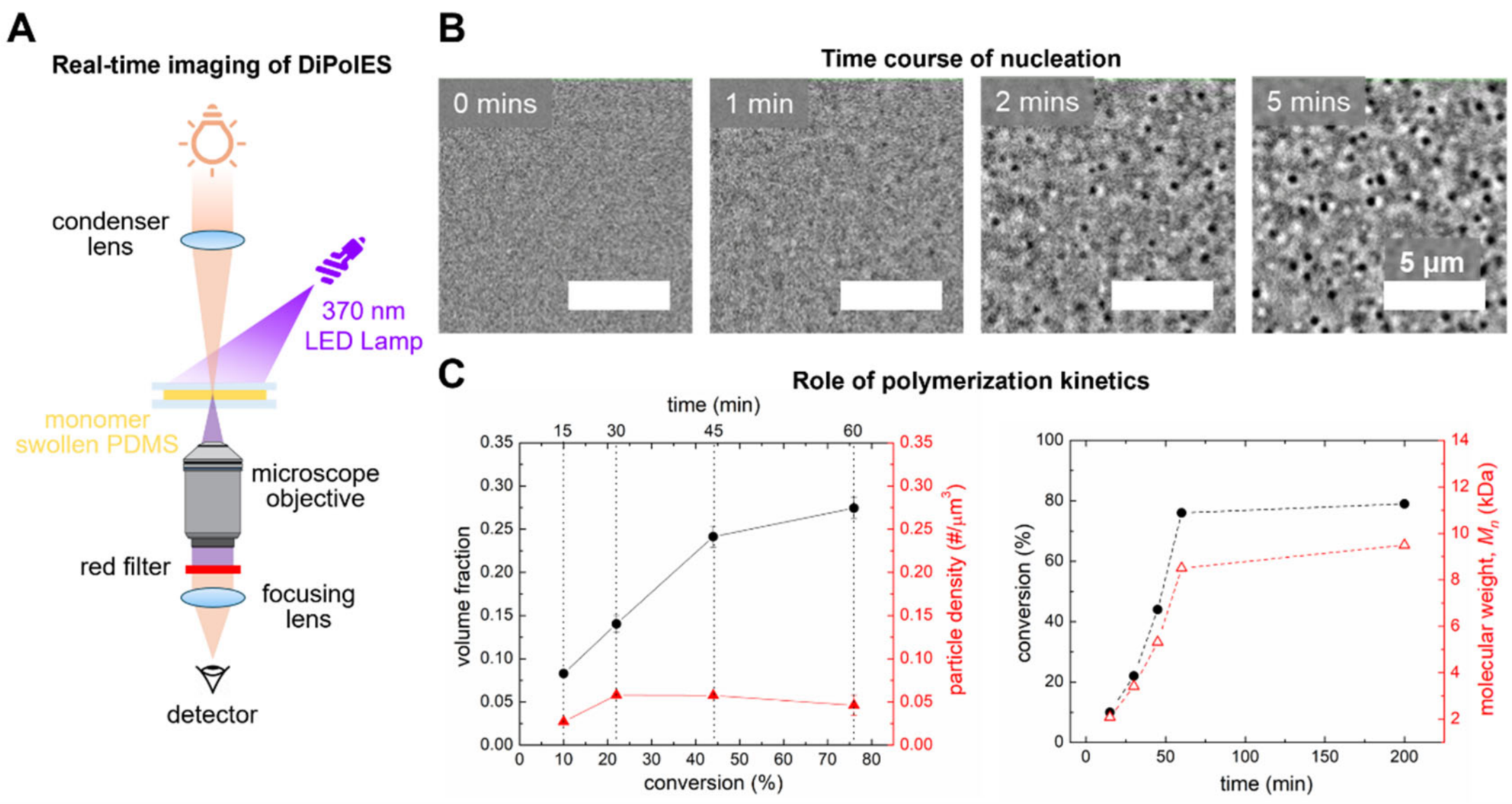

**Figure 2.** The mechanism of particle growth is revealed through real-time imaging and analysis. (A) Schematic of brightfield microscopy for real-time DiPolES imaging. (B) Micrographs from early timepoints show distinct PMMA particle formation within two minutes of light exposure. (C) After nucleation, particle density stays constant with time, while particle volume increases. During this growth phase, conversion of MMA and $M_n$ of PMMA increase linearly with time until plateauing after 1 h.

## 2.3 Composition of particles characterized by EDS and Raman microscopy suggest particles form by cavitation

Because DiPolES employs a solid elastomer as the solvent, the fate of the elastomeric network during particle formation is not immediately evident. Previous studies of crystal growth in gels report partial incorporation of the polymer network into the growing crystal,[39] while experiments on liquid phase separation in elastomers found no detectable network material within the droplets.[29,30] To probe this question, we performed volumetric and surface characterization on a DiPolES sample made at high MMA loading ($C_{liq}$ = 40 %) in a PDMS elastomeric solvent with a 100 kPa modulus. Upon polymerization to completion, we suspended slices of this sample in a 100 μM rhodamine/ethanol solution. Leveraging the preferential partitioning of rhodamine into the PMMA phase, we acquired fluorescent Z-stack images *via* confocal fluorescence microscopy (CFM). Volumetric 3D reconstruction of CFM Z-stacks and complementary scanning electron microscopy (SEM) images reveal that DiPolES results in dispersed particles embedded within elastomeric solvent (Fig. 3A).

To verify this mechanism of network cavitation during DiPolES, we performed energy-dispersive X-ray spectroscopy (EDS) and confocal Raman imaging to characterize particle composition (see *Supporting Information* for details). Using SEM, we first acquired a surface profile of a sliced sample to overlay the measured elemental X-ray spectra (Fig. 3B, right). EDS spectra are collected at an accelerating volage of 5keV to create an elemental map of silicon (1.74 keV; green channel) and carbon (0.277 keV; red channel). SEM-EDS overlay reveals a markedly reduced Si signal (green channel) within the particles (Fig. 3B, right). Complementary confocal Raman spectra on a similar area scan containing particles confirms this result (Fig. 3C). Using the characteristic peak intensity for PDMS and PMMA, identified at 710 $cm^{-1}$ and 1730 $cm^{-1}$ respectively,[40,41] we quantify the relative abundance of silicone-related vibrational modes inside the particles[42] (Fig. 3C, left). The quantified PDMS signal within the particle is comparable to the detection threshold, possibly due to out-of-plane signals (Fig. 3C, right). Together, these findings indicate that the PDMS network remains largely excluded from the growing particles. Analogous to cavitation under tensile loading, growing particles create enormous deformations of the elastomer.[29] Thus, we expect there to be large residual strains in the elastomer following polymerization, which could impact particle growth.[43]

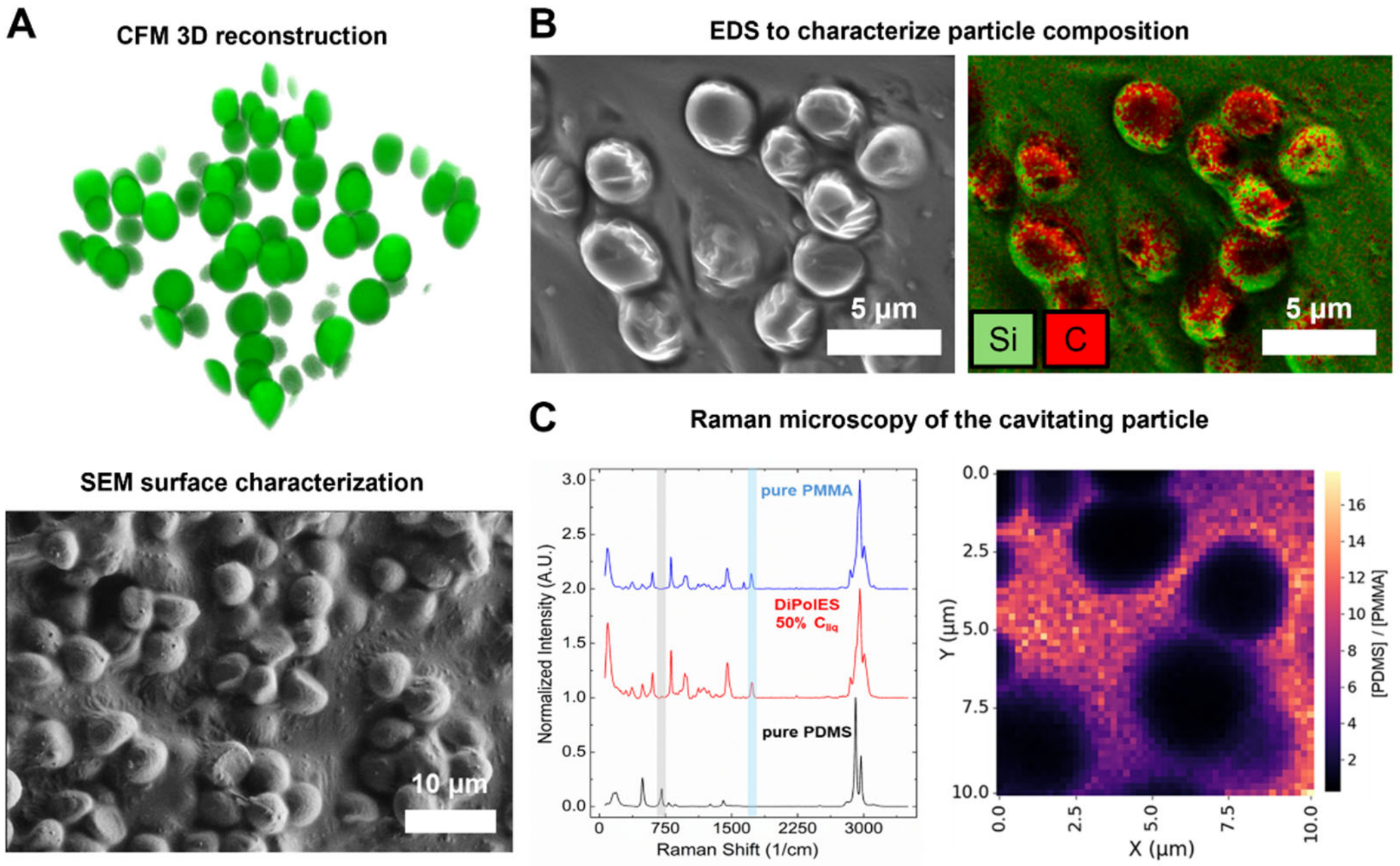

**Figure 3.** Particle composition through microscopic and spectroscopic techniques. (A) 3D reconstruction of particles (top) from samples with a E = 100 kPa and a monomer loading of 40 wt% imaged using confocal fluorescence microscopy. An SEM micrograph (bottom) showing particle and matrix topology from a cross-sectional slice taken in the bulk of a sample. (B) SEM micrograph and corresponding elemental map from EDS, showing particles void of Si signal embedded in the solvent matrix rich in both C and Si signal. (C) Characteristic Raman spectra (left) for pure PDMS (black), pure PMMA (blue), and the interior of a PMMA particle in a DiPolES sample (red). Highlighted are the spectral regions of interest that are unique to PDMS (~710 cm$^{-1}$; grey shading) and PMMA (~1730 cm$^{-1}$; blue shading). Spectral heat map of relative PDMS abundance (right) from a 10x10 µm2 area scan shows the PDMS network is excluded from the particle, suggesting network cavitation during DiPolES.

## 2.4 Effect of network elasticity and monomer loading on particle morphology and volume fraction

To probe the impact of network elasticity on particle growth, we conducted photocontrolled polymerizations of MMA inside PDMS of varying $E$ over a range of $C_{liq}$ values. Upon systematically varying these parameters, we characterized particle morphology using the previously mentioned CFM methods. Surprisingly, we found that the stiffness of the elastomer has little effect on average particle size. For example, at a fixed $C_{liq}$ of 20%, PMMA particles have an average diameter of 0.8 µm, irrespective of PDMS network stiffness, from 10 to 650 kPa (Fig.

4A). This contrasts with recent studies on thermally driven phase separation of liquids in the same PDMS networks, wherein smaller droplets were observed in stiffer networks.[28,30] Here, particle size increases (Fig. 4A, Fig. S8A), while the particle number density decreases (Fig. S6B, S8B) with monomer loading. This observation further confirms that DiPolES maintains the mechanistic features of conventional dispersion polymerization; the particle size increases with monomer concentration and is observed to be inversely correlated with its number density[cite]. In stiff samples, we observe a morphological transition from spheroids (blue circles) to clusters (orange stars) at high monomer loadings (Fig. S6C). Interestingly, the $C_{liq}$ at which we observe clusters drops with network stiffness, from 50% at 100 kPa to 30% at 650 kPa, as shown in the morphological phase diagram of Figure 4A. Cluster morphologies could reflect underlying network heterogeneities[44] or a mechanical transition from cavitation to fracture.[29,45] In the following analysis, we focus our attention on spheroidal particles.

The size distributions of the phase-separated PMMA domains are quantified from the analysis of the acquired CFM images (see *Supporting Information*). At 100 kPa stiffness, particle sizes increase from 0.85 μm to 3.1 μm as $C_{liq}$ increases from 10% to 30% (Fig. 4B). This ability to tune particle size using just monomer loading is an advantageous feature unique to DiPolES. In contrast, particle size control in a conventional dispersion technique relies on a complex interplay of stabilizer concentration, solvent quality, and initiator kinetics.[12–14] DiPolES particle size distributions remain unimodal for monomer loadings up to 20% with polydispersity index ($\sigma/\mu$) ranging from 1 to 9%. We attribute the observed homogenous size distribution to the controlled photoiniferter mechanism employed during DiPolES (Fig. S4, S8C) , as an analogous DiPolES of MMA free radical polymerization results in irregular particles with heterogeneous size distribution (Fig. S9). In addition, the volume fraction of particles formed in the elastomeric solvent correlates linearly with the amount of monomer loaded during DiPolES (Fig. 4C). Thus, we can achieve a particle yield as high as 50% of the total sample volume without modifying the elastomeric solvent.

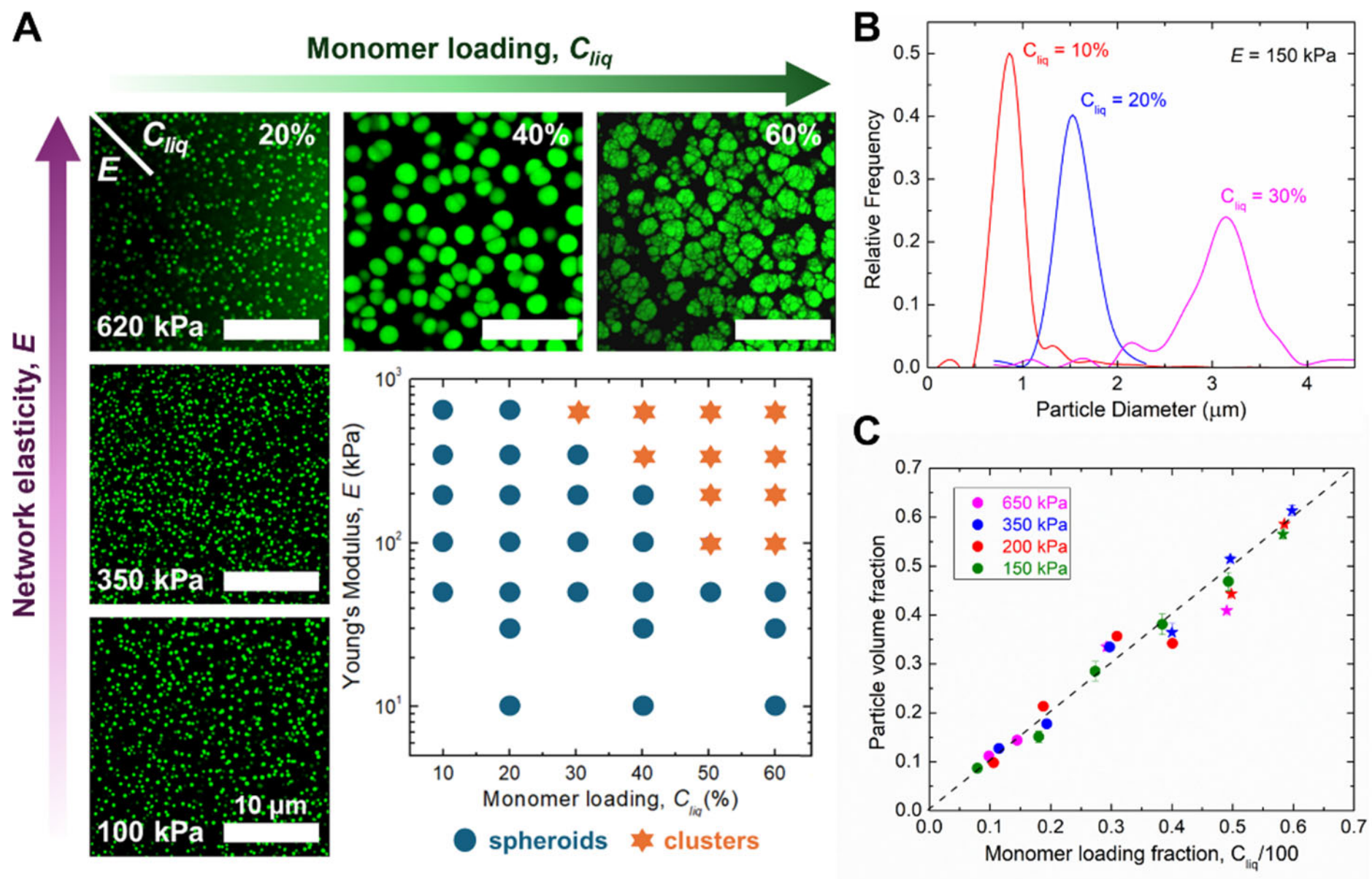

**Figure 4.** Particle morphology and production volume are tuned by the monomer loading in the elastomer. (A) Morphological phase diagram constructed by varying the monomer loading ($C_{liq}$) and network elasticity (E) and characterizing particle morphology as either spheroids (blue circles) or clusters (orange stars). (B) Increasing $C_{liq}$ yields particles with greater diameters while maintaining a narrow size distribution. (C) Particle volume fraction for both spheroids (circles) and clusters (stars) scales with $C_{liq}$, yielding spheroidal particles content up to 50% of the total sample volume.

Overall, these results demonstrate that DiPolES provides a simple yet highly tunable platform for fabricating uniform particles at high yields, with the monomer loading and solvent stiffness serving as effective control parameters for particle size and morphology, respectively. By coupling controlled polymerization with mechanical stabilization of the elastomeric solvent, DiPolES enables access to a broad morphological phase space as opposed to conventional techniques. For example, hybrid elastomeric materials comprising PDMS/PMMA interpenetrating polymer networks (IPNs), typically yield heterogeneous structures with limited control over feature size and morphology.[24,25] By leveraging the chemical-mechanical coupling, DiPolES extends this concept to broader length scales, enabling a robust approach to engineer multiphase polymeric composites. The mechanically programmable structure morphology observed in DiPolES requires systematic characterization of network architecture, which we will explore in future studies.

### 2.5 DiPolES produces ellipsoidal particles in anisotropically strained elastomers

Given the impact of solvent elasticity on particle morphology, we decided to probe the mechanical stress on elastic matrix and the resulting anisotropy of the particle formed. Without external stress, interfacial tension and isotropic network forces of the elastomer favor spherical particles.[45] However, recent work on condensing liquid particles has shown that particle symmetry can easily be broken by stretching the swollen elastomer along one axis.[29] We hypothesize that mechanical deformation of the elastomeric solvent during DiPolES could similarly lead to *in-situ* growth of anisotropic structures.[29,45] To test this, we performed DiPolES under uniaxial strain by stretching PDMS ($E$ = 30 kPa) swollen with MMA/CTA ($C_{liq}$ = 40%) using a custom-built mechanical stretcher, followed by photopolymerization under the same conditions as previously described (Fig. 5A). Under 50% strain, we observed that the resulting PMMA particles adopted ellipsoidal morphologies, with their long axes aligned along the direction of applied strain (Fig. 5B). Remarkably, the ellipticity of these particles is preserved even after the PDMS matrix is released from its stretched state, due to the glassy nature of the PMMA phase that kinetically traps the deformed shape. When a crosslinking monomer is included, these ellipsoidal particles are resistant to solvent treatment and can be extracted from the matrix following previously mentioned strategy of depolymerizing the PDMS (see *Supporting Information* for details) (Fig. 5C).[36] Typical strategies to fabricate such elliptical particles rely on high temperature post-processing of preformed spherical particles[46] or two-step seeded polymerization,[47] which are energy-intensive, low-throughput, and challenging to control at scale. In contrast, here we show that the combination of elastic confinement and controlled mechanical strain in DiPolES provides a new means of directing particle shape; a single-step approach for fabricating elliptical particles, which is otherwise inaccessible with liquid-state dispersion polymerization or PIPS systems.

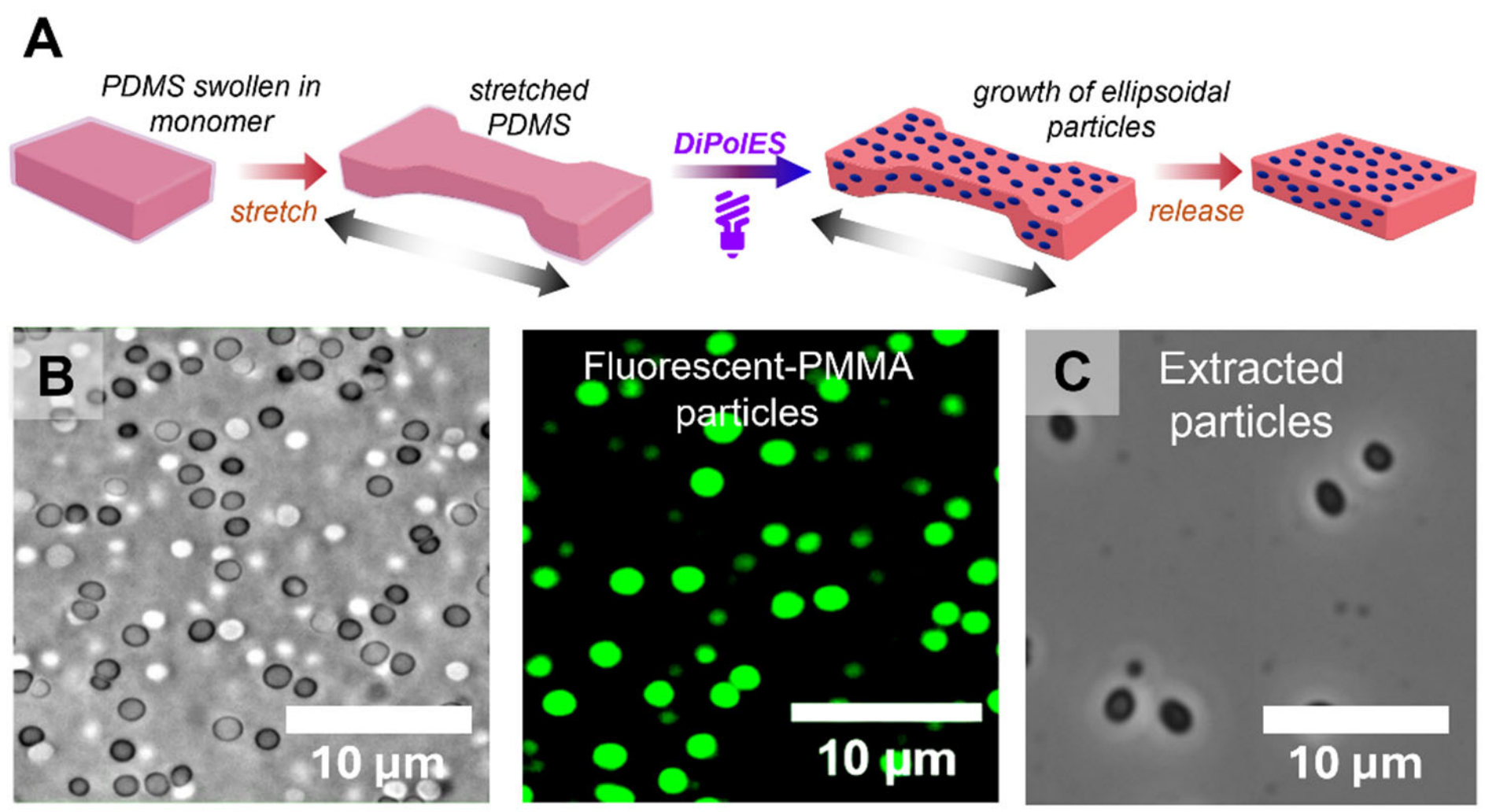

**Figure 5.** Anisotropic DiPolES. (A) Performing DiPolES on stretched samples produces ellipsoidal particles. (B) Brightfield (left) and fluorescent (middle) micrographs show ellipsoidal particles formed under anisotropic DiPolES. (C) Following depolymerization of PDMS matrix, brightfield microscopy reveals that particles maintain their ellipsoidal morphology.

## 3. Conclusion

In this work, we introduce DiPolES (Dispersion Polymerization in an Elastomeric Solvent) as a versatile and scalable strategy for microscale particle fabrication. In this system, photoiniferter polymerization results in simultaneous particle nucleation at low monomer conversions, while matrix elasticity confines particle growth for the remaining controlled polymer propagation, resulting in uniform size distributions. By varying the monomer loading and elastomer stiffness, particle size and morphologies can be tuned, respectively. Notably, we demonstrate that stretching the elastomer is a facile, efficient route to generate highly desirable ellipsoidal particles, which can be extracted following elastomer depolymerization. We anticipate that DiPolES will be effective for a wide range of monomer–elastomer combinations, polymerization mechanisms (e.g. ROMP, cationic, thiol-ene, etc.), and stimulus-responsive initiation strategies. Extending this platform to monomers with optical, electronic, or stimuli-responsive properties may yield novel materials with functional emergent behavior. Ultimately, DiPolES represents a powerful paradigm for fabricating multiphase materials that bridge the gap between particle synthesis and composite engineering.

ASSOCIATED CONTENT

**Supporting Information**.

The following file is available free of charge.

- Supporting Information: General Reagent Information, General Analytical Information, General Synthetic Procedure (S1-S7) (PDF)

AUTHOR INFORMATION

**Corresponding Author**

*Email: erd43@cornell.edu

**Present Addresses**

†H.K.: Institute of Materials, EPFL, Lausanne, Switzerland

**Author Contributions**

The manuscript was written through contributions of all authors. All authors have given approval to the final version of the manuscript.

**Funding Sources**

Graduate research fellowships from the National Science Foundation to TB and RJD (DGE – 2139899); Research grants to ERD and BPF from the Department of Energy (DE-SC0026177) and Sandia National Laboratories (LDRD 240531).

ACKNOWLEDGMENT

We acknowledge support for the development of reaction schemes and chemical analysis through graduate research fellowships from the National Science Foundation to TEB and RJD (DGE – 2139899). SD's morphological measurements and phase diagrams were supported by Department of Energy Office of Science, Office of Basic Energy Sciences Biomolecular Materials program under Award Number (DE-SC0026177). EEO's characterization of cavitation was supported by Sandia National Laboratories (LDRD 240531). This work made use of the Cornell Center for Materials Research shared instrumentation facility. We thank Robert Style for feedback on the manuscript.

REFERENCES

(1) Whitesides, G. M.; Grzybowski, B. Self-Assembly at All Scales. *Science* **2002**, *295* (5564), 2418–2421. https://doi.org/10.1126/science.1070821.

(2) James, P. F. Liquid-Phase Separation in Glass-Forming Systems. *J. Mater. Sci.* **1975**, *10* (10), 1802–1825. https://doi.org/10.1007/BF00554944.

(3) Lee, K.-W. D.; Chan, P. K.; Feng, X. Morphology Development and Characterization of the Phase-Separated Structure Resulting from the Thermal-Induced Phase Separation Phenomenon in Polymer Solutions under a Temperature Gradient. *Chem. Eng. Sci.* **2004**, *59* (7), 1491–1504. https://doi.org/10.1016/j.ces.2003.12.025.

(4) Zhao, J.; Luo, G.; Wu, J.; Xia, H. Preparation of Microporous Silicone Rubber Membrane with Tunable Pore Size via Solvent Evaporation-Induced Phase Separation. *ACS Appl. Mater. Interfaces* **2013**, *5* (6), 2040–2046. https://doi.org/10.1021/am302929c.

(5) Boots, H. M. J.; Kloosterboer, J. G.; Serbutoviez, C.; Touwslager, F. J. Polymerization-Induced Phase Separation. 1. Conversion−Phase Diagrams. *Macromolecules* **1996**, *29* (24), 7683–7689. https://doi.org/10.1021/ma960292h.

(6) Liu, Y. Polymerization-Induced Phase Separation and Resulting Thermomechanical Properties of Thermosetting/Reactive Nonlinear Polymer Blends: A Review. *J. Appl. Polym. Sci.* **2013**, *127* (5), 3279–3292. https://doi.org/10.1002/app.38721.

(7) Szczepanski, C. R.; Pfeifer, C. S.; Stansbury, J. W. A New Approach to Network Heterogeneity: Polymerization Induced Phase Separation in Photo-Initiated, Free-Radical Methacrylic Systems. *Polymer* **2012**, *53* (21), 4694–4701. https://doi.org/10.1016/j.polymer.2012.08.010.

(8) Sperling, L. H.; Hu, R. Interpenetrating Polymer Networks. In *Polymer Blends Handbook*; Springer, Dordrecht, 2003; pp 417–447. https://doi.org/10.1007/0-306-48244-4_6.

(9) Lohani, A.; Singh, G.; Bhattacharya, S. S.; Verma, A. Interpenetrating Polymer Networks as Innovative Drug Delivery Systems. *J. Drug Deliv.* **2014**, *2014* (1), 583612. https://doi.org/10.1155/2014/583612.

(10) Farooq, U.; Teuwen, J.; Dransfeld, C. Toughening of Epoxy Systems with Interpenetrating Polymer Network (IPN): A Review. *Polymers* **2020**, *12* (9), 1908. https://doi.org/10.3390/polym12091908.

(11) Sperling, L. H. Interpenetrating Polymer Networks. In *Encyclopedia of Polymer Science and Technology*; John Wiley & Sons, Ltd, 2004. https://doi.org/10.1002/0471440264.pst170.

(12) Arshady, R. Suspension, Emulsion, and Dispersion Polymerization: A Methodological Survey. *Colloid Polym. Sci.* **1992**, *270* (8), 717–732. https://doi.org/10.1007/BF00776142.

(13) Kawaguchi, S.; Ito, K. Dispersion Polymerization. In *Polymer Particles*; Okubo, M., Ed.; Springer: Berlin, Heidelberg, 2005; pp 299–328. https://doi.org/10.1007/b100118.

(14) Sudol, E. D. Dispersion Polymerization. In *Polymeric Dispersions: Principles and Applications*; Asua, J. M., Ed.; Springer Netherlands: Dordrecht, 1997; pp 141–153. https://doi.org/10.1007/978-94-011-5512-0_10.

(15) Kawaguchi, H. Functional Polymer Microspheres. *Prog. Polym. Sci.* **2000**, *25* (8), 1171–1210. https://doi.org/10.1016/S0079-6700(00)00024-1.

(16) Oh, J. K.; Drumright, R.; Siegwart, D. J.; Matyjaszewski, K. The Development of Microgels/Nanogels for Drug Delivery Applications. *Prog. Polym. Sci.* **2008**, *33* (4), 448–477. https://doi.org/10.1016/j.progpolymsci.2008.01.002.

(17) Saralidze, K.; Koole, L. H.; Knetsch, M. L. W. Polymeric Microspheres for Medical Applications. *Materials* **2010**, *3* (6), 3537–3564. https://doi.org/10.3390/ma3063537.

(18) Pulingam, T.; Foroozandeh, P.; Chuah, J.-A.; Sudesh, K. Exploring Various Techniques for the Chemical and Biological Synthesis of Polymeric Nanoparticles. *Nanomaterials* **2022**, *12* (3), 576. https://doi.org/10.3390/nano12030576.

(19) Shagymgereyeva, S.; Sarsenbekuly, B.; Kang, W.; Yang, H.; Turtabayev, S. Advances of Polymer Microspheres and Its Applications for Enhanced Oil Recovery. *Colloids Surf. B Biointerfaces* **2024**, *233*, 113622. https://doi.org/10.1016/j.colsurfb.2023.113622.

(20) Zielińska, A.; Carreiró, F.; Oliveira, A. M.; Neves, A.; Pires, B.; Venkatesh, D. N.; Durazzo, A.; Lucarini, M.; Eder, P.; Silva, A. M.; Santini, A.; Souto, E. B. Polymeric Nanoparticles: Production, Characterization, Toxicology and Ecotoxicology. *Molecules* **2020**, *25* (16), 3731. https://doi.org/10.3390/molecules25163731.

(21) Paine, A. J. Dispersion Polymerization of Styrene in Polar Solvents. 7. A Simple Mechanistic Model to Predict Particle Size. *Macromolecules* **1990**, *23* (12), 3109–3117. https://doi.org/10.1021/ma00214a013.

(22) Sicher, A.; Ganz, R.; Menzel, A.; Messmer, D.; Panzarasa, G.; Feofilova, M.; O. Prum, R.; W. Style, R.; Saranathan, V.; M. Rossi, R.; R. Dufresne, E. Structural Color from Solid-State Polymerization-Induced Phase Separation. *Soft Matter* **2021**, *17* (23), 5772–5779. https://doi.org/10.1039/D1SM00210D.

(23) Sicher, A.; Whitfield, R.; Ilavsky, J.; Saranathan, V.; Anastasaki, A.; Dufresne, E. R. ATRP Enhances Structural Correlations In Polymerization-Induced Phase Separation. *Angew. Chem. Int. Ed.* **2023**, *62* (16), e202217683. https://doi.org/10.1002/anie.202217683.

(24) Silvaroli, A. J.; Heyl, T. R.; Qiang, Z.; Beebe, J. M.; Ahn, D.; Mangold, S.; Shull, K. R.; Wang, M. Tough, Transparent, Photocurable Hybrid Elastomers. *ACS Appl. Mater. Interfaces* **2020**, *12* (39), 44125–44136. https://doi.org/10.1021/acsami.0c11643.

(25) Beebe, J. M.; Ahn, D.; Eldred, D. V.; Fielitz, A. J.; Heyl, T. R.; Lee, M.; Mangold, S.; Pearce, E. Z.; Reinhardt, C. W.; Roggenbuck, C.; Scherzer, J. M.; Shull, K. R.; Silvaroli, A. J.; Tan, Y.-J.; Wang, M. Photocured Simultaneous and Sequential PDMS/PMMA Interpenetrating Polymer Networks. *Macromolecules* **2022**, *55* (13), 5826–5839. https://doi.org/10.1021/acs.macromol.2c00425.

(26) Lipatov, Y. S.; Alekseeva, T. T. Phase-Separated Interpenetrating Polymer Networks. In *Phase-Separated Interpenetrating Polymer Networks*; Springer: Berlin, Heidelberg, 2007; pp 1–227. https://doi.org/10.1007/12_2007_116.

(27) Heyl, T. R.; Beebe, J. M.; Silvaroli, A. J.; Perce, A.; Ahn, D.; Mangold, S.; Mazure, V.; Shull, K. R.; Wang, M. In Situ Investigations of Microstructure Formation in Interpenetrating Polymer Networks. *Macromolecules* **2024**, *57* (5), 1950–1961. https://doi.org/10.1021/acs.macromol.3c02097.

(28) Style, R. W.; Sai, T.; Fanelli, N.; Ijavi, M.; Smith-Mannschott, K.; Xu, Q.; Wilen, L. A.; Dufresne, E. R. Liquid-Liquid Phase Separation in an Elastic Network. *Phys. Rev. X* **2018**, *8* (1), 011028. https://doi.org/10.1103/PhysRevX.8.011028.

(29) Kim, J. Y.; Liu, Z.; Weon, B. M.; Cohen, T.; Hui, C.-Y.; Dufresne, E. R.; Style, R. W. Extreme Cavity Expansion in Soft Solids: Damage without Fracture. *Sci. Adv.* **2020**, *6* (13), eaaz0418. https://doi.org/10.1126/sciadv.aaz0418.

(30) Fernández-Rico, C.; Schreiber, S.; Oudich, H.; Lorenz, C.; Sicher, A.; Sai, T.; Bauernfeind, V.; Heyden, S.; Carrara, P.; Lorenzis, L. D.; Style, R. W.; Dufresne, E. R. Elastic Microphase Separation Produces Robust Bicontinuous Materials. *Nat. Mater.* **2024**, *23* (1), 124–130. https://doi.org/10.1038/s41563-023-01703-0.

(31) Rosowski, K. A.; Vidal-Henriquez, E.; Zwicker, D.; Style, R. W.; Dufresne, E. R. Elastic Stresses Reverse Ostwald Ripening. *Soft Matter* **2020**, *16* (25), 5892–5897. https://doi.org/10.1039/D0SM00628A.

(32) Rosowski, K. A.; Sai, T.; Vidal-Henriquez, E.; Zwicker, D.; Style, R. W.; Dufresne, E. R. Elastic Ripening and Inhibition of Liquid–Liquid Phase Separation. *Nat. Phys.* **2020**, *16* (4), 422–425. https://doi.org/10.1038/s41567-019-0767-2.

(33) Zhang, Z.; Jiang, W.; Wang, Y.; Wu, Y.; Hou, X. Synthesis of Monodispersed Polyacrylonitrile Microspheres by Dispersion Polymerization. *J. Appl. Polym. Sci.* **2012**, *125* (5), 4142–4148. https://doi.org/10.1002/app.36407.

(34) Shiho, H.; DeSimone, J. M. Dispersion Polymerization of Acrylonitrile in Supercritical Carbon Dioxide. *Macromolecules* **2000**, *33* (5), 1565–1569. https://doi.org/10.1021/ma990737c.

(35) Fujii, S.; Kameyama, S.; Armes, S. P.; Dupin, D.; Suzaki, M.; Nakamura, Y. pH-Responsive Liquid Marbles Stabilized with Poly(2-Vinylpyridine) Particles. *Soft Matter* **2010**, *6* (3), 635–640. https://doi.org/10.1039/B914997J.

(36) Warner, M. J.; Kopatz, J. W.; Schafer, D. P.; Kustas, J.; Sawyer, P. S.; Grillet, A. M.; Jones, B. H.; Ghosh, K. A Robust Depolymerization Route for Polysiloxanes. *Chem. Commun.* **2024**, *60* (9), 1188–1191. https://doi.org/10.1039/D3CC05509D.

(37) Ober, C. K.; Lok, K. P.; Hair, M. L. Monodispersed, Micron-Sized Polystyrene Particles by Dispersion Polymerization. *J. Polym. Sci. Polym. Lett. Ed.* **1985**, *23* (2), 103–108. https://doi.org/10.1002/pol.1985.130230209.

(38) György, C.; Verity, C.; Neal, T. J.; Rymaruk, M. J.; Cornel, E. J.; Smith, T.; Growney, D. J.; Armes, S. P. RAFT Dispersion Polymerization of Methyl Methacrylate in Mineral Oil: High Glass Transition Temperature of the Core-Forming Block Constrains the Evolution of Copolymer Morphology. *Macromolecules* **2021**, *54* (20), 9496–9509. https://doi.org/10.1021/acs.macromol.1c01528.

(39) Li, H.; Xin, H. L.; Muller, D. A.; Estroff, L. A. Visualizing the 3D Internal Structure of Calcite Single Crystals Grown in Agarose Hydrogels. *Science* **2009**, *326* (5957), 1244–1247. https://doi.org/10.1126/science.1178583.

(40) Willis, H. A.; Zichy, V. J. I.; Hendra, P. J. The Laser-Raman and Infra-Red Spectra of Poly(Methyl Methacrylate). *Polymer* **1969**, *10*, 737–746. https://doi.org/10.1016/0032-3861(69)90101-3.

(41) Cai, D.; Neyer, A.; Kuckuk, R.; Heise, H. M. Raman, Mid-Infrared, near-Infrared and Ultraviolet–Visible Spectroscopy of PDMS Silicone Rubber for Characterization of Polymer Optical Waveguide Materials. *J. Mol. Struct.* **2010**, *976* (1), 274–281. https://doi.org/10.1016/j.molstruc.2010.03.054.

(42) Pelletier, M. J. Quantitative Analysis Using Raman Spectrometry. *Appl. Spectrosc.* **2003**, *57* (1), 20A-42A. https://doi.org/10.1366/000370203321165133.

(43) Fernández-Rico, C.; Style, R. W.; Heyden, S.; Wang, S.; Olmsted, P. D.; Dufresne, E. R. Thermodynamics of Microphase Separation in a Swollen, Strain-Stiffening Polymer Network. *Soft Matter* **2026**, *22* (2), 330–342. https://doi.org/10.1039/D5SM00594A.

(44) Qiang, Y.; Luo, C.; Zwicker, D. Nonlocal Elasticity Yields Equilibrium Patterns in Phase Separating Systems. *Phys. Rev. X* **2024**, *14* (2), 021009. https://doi.org/10.1103/PhysRevX.14.021009.

(45) Raayai-Ardakani, S.; Earl, D. R.; Cohen, T. The Intimate Relationship between Cavitation and Fracture. *Soft Matter* **2019**, *15* (25), 4999–5005. https://doi.org/10.1039/C9SM00570F.

(46) Ho, C. C.; Keller, A.; Odell, J. A.; Ottewill, R. H. Preparation of Monodisperse Ellipsoidal Polystyrene Particles. *Colloid Polym. Sci.* **1993**, *271* (5), 469–479. https://doi.org/10.1007/BF00657391.

(47) Sheu, H. R.; El-Aasser, M. S.; Vanderhoff, J. W. Phase Separation in Polystyrene Latex Interpenetrating Polymer Networks. *J. Polym. Sci. Part Polym. Chem.* **1990**, *28* (3), 629–651. https://doi.org/10.1002/pola.1990.080280314.